\documentclass[%
 aps,
 prb,%
 amsmath,amssymb,
preprint,%
superscriptaddress,showpacs]{revtex4-1}

\usepackage{color, graphicx}
\usepackage{dcolumn}
\usepackage{bm}
\usepackage{amssymb}
\usepackage{latexsym}
\usepackage{amsfonts}
\usepackage{amsmath}
\usepackage{multirow}
\usepackage{ifthen}

\begin{document}


\title{Level repulsion and evanescent waves in sonic crystals}

\author{V. Romero-Garc\'ia}
 \affiliation{Instituto de Investigaci\'on para la Gesti\'on Integrada de zonas Costeras, Universidad Polit\'ecnica de Valencia, Paraninf 1, 46730, Gandia, Spain}
 
 \author{J.O. Vasseur}
  \email{jerome.vasseur@univ-lille1.fr}
\author{A.C. Hladky-Hennion} 
\affiliation{Institut d'Electronique, de Micro\'electronique et de Nanotechnologie,  UMR CNRS 8520, Cit\'e Scientifique, F-59652 Villeneuve d'Ascq C\'edex, France}
 
\author{L.M. Garcia-Raffi}
 \affiliation{Instituto Universitario de Matem\'atica Pura y Aplicada, Universidad Polit\'ecnica de Valencia, Camino de Vera s/n, 46022, Valencia, Spain}
 
 \author{J.V. S\'anchez-P\'erez}
 \affiliation{Centro de Tecnolog\'ias F\'isicas: Ac\'ustica, Materiales y Astrof\'isica, Universidad Polit\'ecnica de Valencia, Camino de Vera s/n, 46022, Valencia, Spain}


\begin{abstract}
This work theoretically and experimentally reports the evanescent connections between propagating bands in periodic acoustic materials. The complex band structures obtained by solving for the $k(\omega)$ problem reveal a complete interpretation of the propagation properties of these systems. The prediction of evanescent modes, non predicted by classical $\omega(\vec{k})$ methods, is of interest for the understanding of these propagation properties. Complex band structures provide an interpretation of the evanescent coupling and the level repulsion states showing the possibility to control of evanescent waves in periodic materials.
\end{abstract}


\maketitle


Propagation of waves in periodic structures is characterized by both propagating and evanescent waves. On one hand, the control of the propagation properties of periodic electromagnetic \cite{Yablonovitch87, John87, joannopoulos08} and elastic \cite{Kushwaha93, Martinez95, Sanchez98, kafesaki99} materials has revealed a huge number of physical phenomena in the last decades. On the other hand, the complex band structures\cite{Laude09, Romero10b} reveal the bandgaps as ranges of frequencies where only evanescent modes can be excited in the system. Imaging with super resolution,\cite{Pendry00, Fang05} that is, overcoming the diffraction limit can be obtained by restoring all the evanescent components of a near-field image. This can be achieved by the coupling of the evanescent modes with other mechanism leading to their amplification in order to successfully transport the information carried by the evanescent waves through the system. Recently, the possibility to control the evanescent properties in periodic composites has shown several interesting possibilities in both photonic \cite{Li07, Dong11} and phononic \cite{Sukhovich09, Robillard11, Zhu11} crystals. In this way, the evanescent properties of the periodic systems have been revealed necessary for the design of new acoustic and electromagnetic applications. 

The characterization of evanescent modes in acoustic periodic structures can be obtained from the complex band structures calculated using $k(\omega)$ methods.\cite{Sainidou05} The extended plane wave expansion (EPWE) with supercell approximation \cite{Romero10c} represents one of the possibilities to evaluate the excited evanescent modes in periodic structures with defects. Although our analysis is general, we study here the evanescent modes in two dimensional (2D) sonic crystal waveguides (SCW) because, as we will see later, this system allows us to obtain high precision measurements inside the periodic structure where the evanescent properties appear. In this work we experimentally and theoretically show that the complex band obtained using the EPWE are more general than the classical ones, $\omega(\vec{k})$, revealing a complete picture of the physics of the systems. 
These results have been also evaluated by solving for the scattering problem using Multiple Scattering Theory (MST).\cite{martin06} 

\begin{figure}[hbt]
\begin{center}
\includegraphics[width=59mm,height=69mm,angle=0]{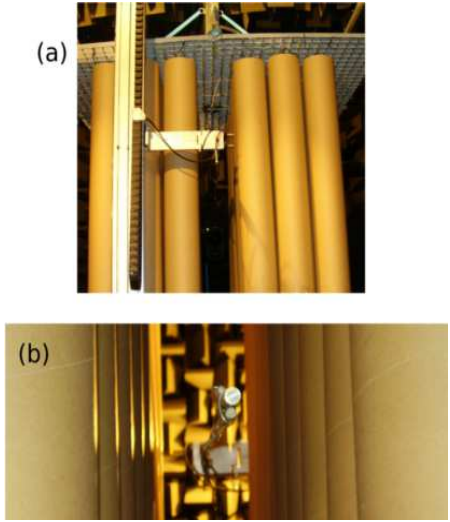}
\end{center} 
\caption{(Color online) (a) Experimental setup. (b) Details of the microphone inside the waveguide.}
 \label{fig:set_up}
\end{figure}

We have performed experiments in an echo-free chamber sized $8\times 6\times 3$ m$^3$ in order to to evaluate all the theoretical predictions.. The SCW has been generated by removing the middle row of a complete square periodic array of rigid cylindrical scatterers embedded in air ($\rho=1.23$ kg/m$^3$, $c_{host}=340$ m/s) with lattice constant $a$, and filling fraction $ff=\pi r^2/a^2=0.71$ (in the experimental setup we use $r=0.09$ m and $a=0.19$ m). To obtain the experimental dependence of the pressure all along the SCW, we measured the pressure field at several points inside the guide. The microphone used was a prepolarized free-field 1/2" Type $4189$ B\&K. The acoustic field has been measured in this work using our 3D Robotized e-Acoustic Measurement System (3DReAMS).\cite{Romero11} Figurea \ref{fig:set_up}a and \ref{fig:set_up}b show detailed pictures of the experimental setup and of the microphone inside the SCW respectively.

\begin{figure}[hbt]
\begin{center}
\includegraphics[width=85mm,height=54mm,angle=0]{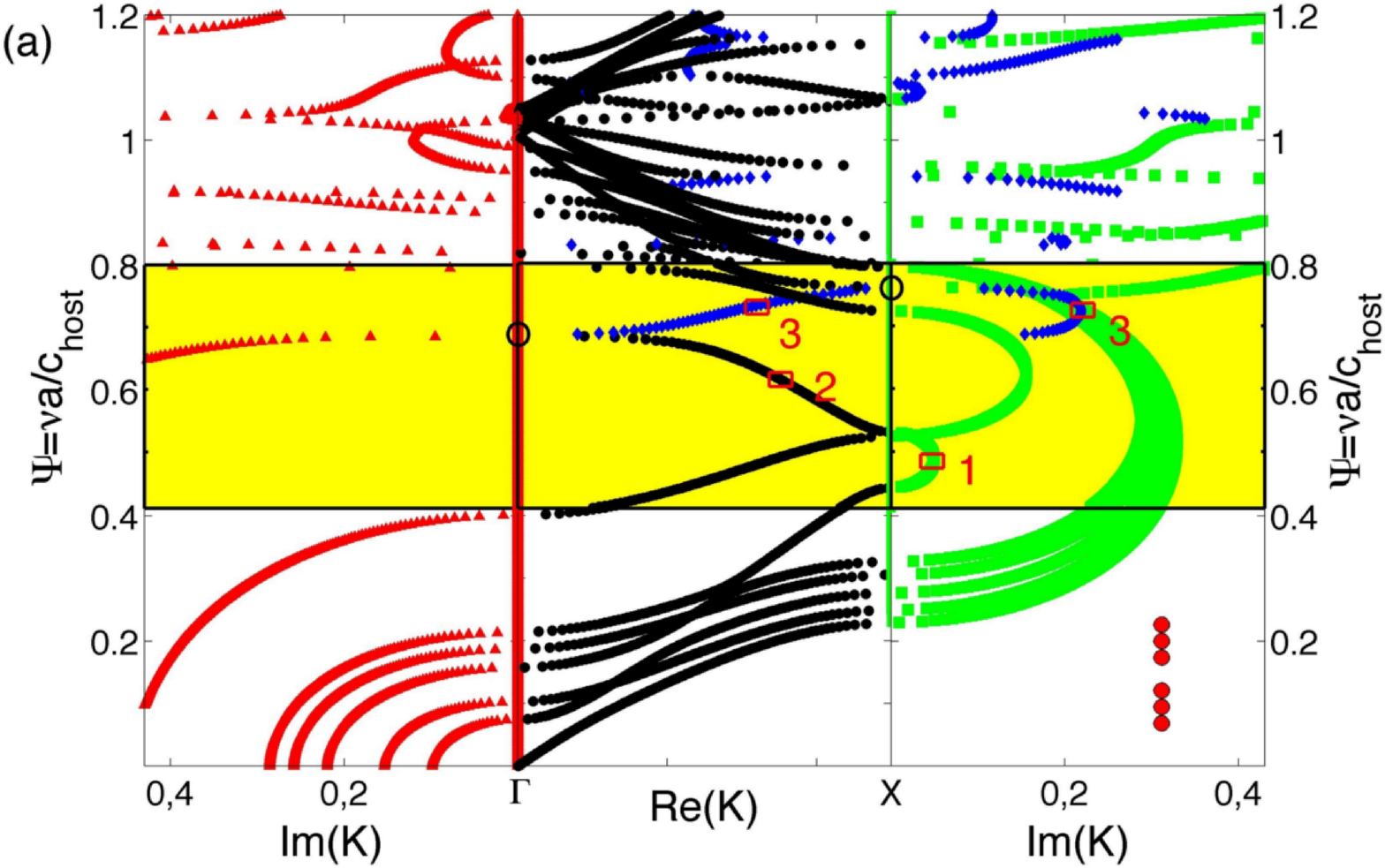}\\
\includegraphics[width=85mm,height=44mm,angle=0]{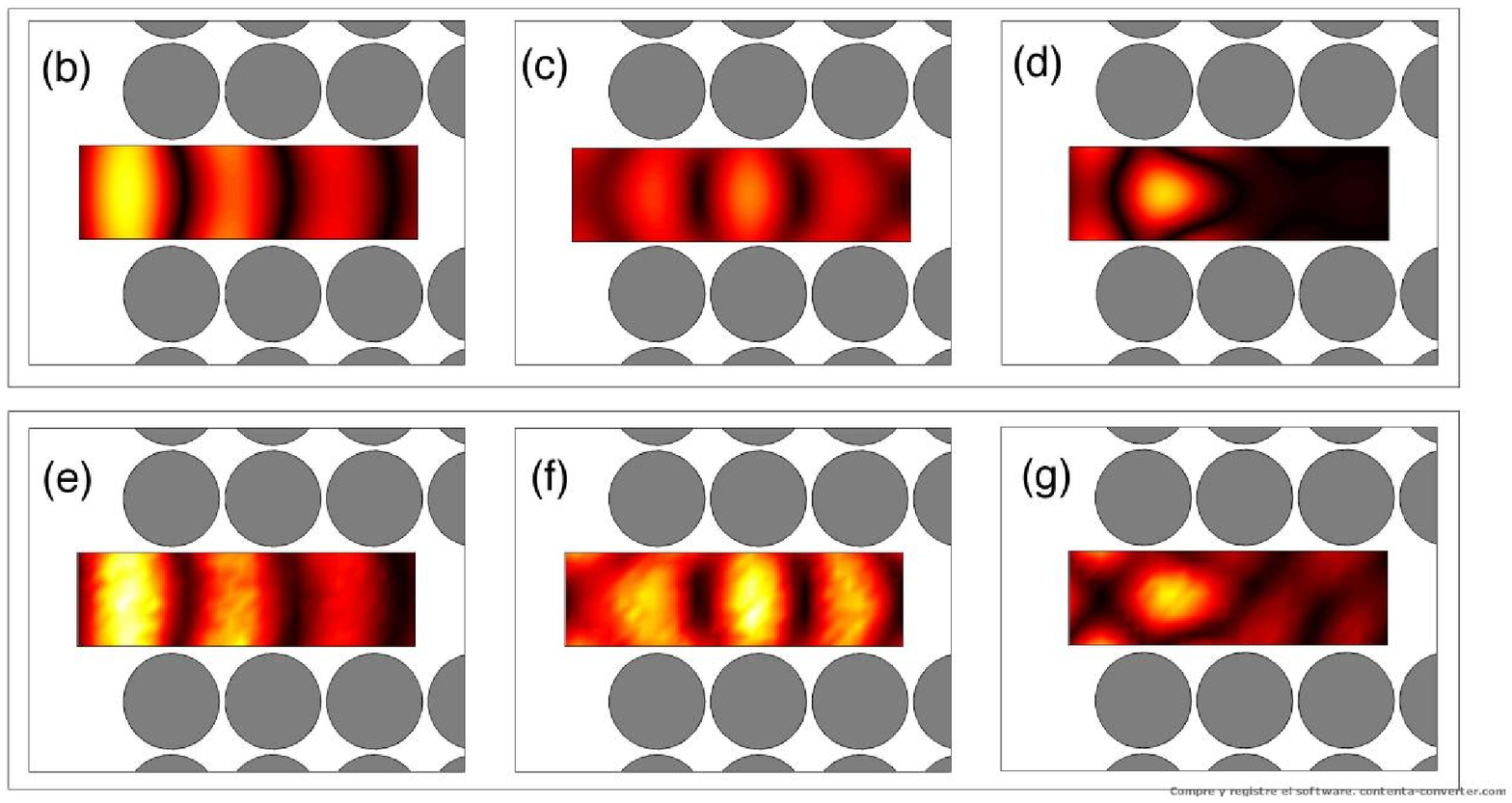}
\end{center} 
\caption{(Color online) Theoretical and experimental analysis of the SCW. (a) Complex band structures for the SCW. 1681 plane waves have been used in the calculation. Reduced magnitudes have been plotted. (b-d) Absolute value of the acoustic field predicted using MST for frequencies (a) $\Psi_1=0.484$, (b) $\Psi_2=0.619$ and (c) $\Psi_3=0.731$. (red squares in (a)). (e), (f) and (g) show the experimental results of the absolute value of the acoustic field for frequencies corresponding to (b), (c) and (d).}
 \label{fig:waveguide}
\end{figure}

Using the classical $\omega(\vec{k})$ methods one can find two kinds of bands, the propagating and the deaf ones,\cite{Sanchez98} and no information can be obtained about the evanescent modes. We show here that the complex band structure reveals additional bands related with evanescent modes, never predicted by the $\omega(\vec{k})$ methods, that drastically change the propagation properties of the system. Figure \ref{fig:waveguide}a shows the complex band structure for a SCW in the $\Gamma$X direction (through the work reduced magnitudes have been used, $\Psi=\nu a/c_{host}$, $K=ka/(2\pi)$). Left and right (respectively, middle) panels represent the imaginary (respectively, real) part of the complex band structures. The classification shown in this work, inspired by the works of Bavencoffe,\cite{Bavencoffe09a} helps to a better understanding of the complex band structures. The convergence of all the calculations has been carefully analyzed selecting the adequate number of plane waves for each case. The modes are classified following the next restrictions: ($i$) The classical band structures correspond to modes characterized by values of $Re(K)$ in the Brillouin zone and $Im(K)=0$. Modes with these properties are shown in this work with black filled circles. ($ii$) The modes characterized by $Im(K)\geq0$ and $Re(K)=0$ are shown with red filled triangles. These modes represent connections between propagating bands at the $\Gamma$ point. ($iii$) The modes characterized by $Im(K)\geq0$ and $Re(K)=1/2$ (respectively, $Re(K)=1/\sqrt{2}$) are shown with green filled squares. These modes represents connections between propagating bands at the $X$ (respectively, M) point. ($iv$) The modes with $Re(K)$ in the first Brillouin zone but with $Im(K)\neq0$ are in this Section in blue filled diamonds. These modes belong to evanescent connecting bands between bands with the same symmetry crossing the Brillouin zone.

\begin{figure*}[hbt]
\begin{center}
\includegraphics[width=170mm,height=75mm,angle=0]{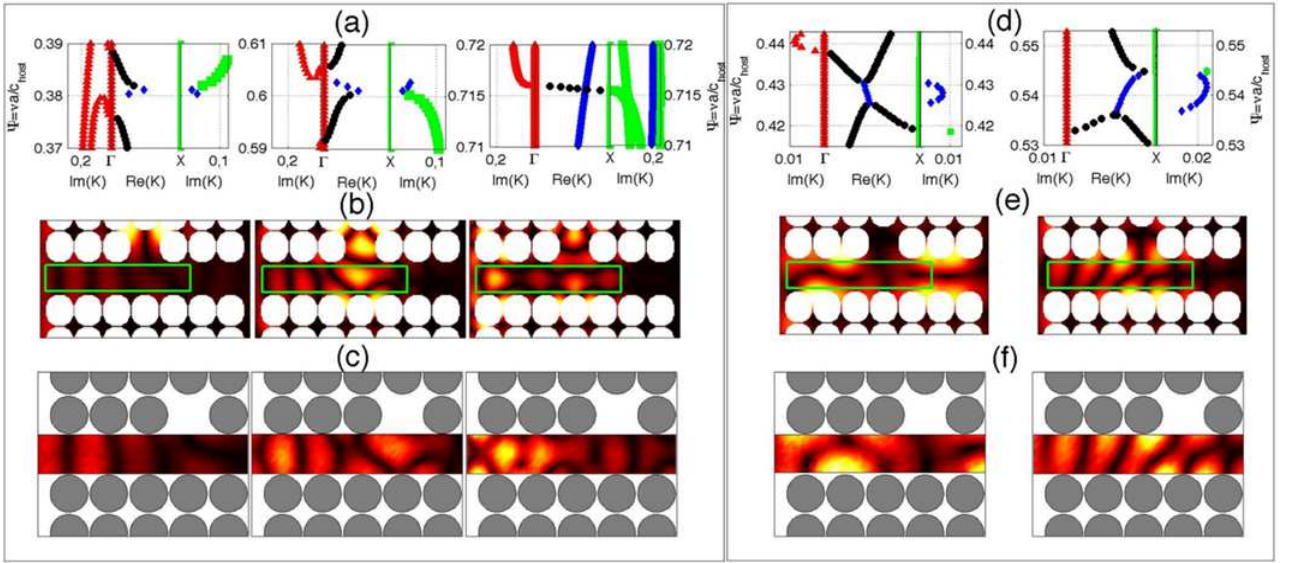}
\end{center} 
\caption{(Color online) Analysis of the stubbed sonic crystal waveguide. Left column: Resonances. The left column shows the results for the first resonance at $\Psi=0.3812$, the middle column shows the ones for the resonance at $\Psi=0.605$ and the right panel for the resonance at $\Psi=0.72$. Right column: Level repulsion. The left panel shows the results for the level repulsion at $\Psi=0.429$ and the right panel the level repulsion at $\Psi=0.544$. (a) and (b) represent the complex band structures calculated using EPWE. (b)-(c) and (e)-(f) represent the MST predictions and the experimental results respectively. We note that  the experimental setup does not allow the measurement of the acoustic field inside the stub. All the maps represent absolute value of the acoustic field. Green rectangles show the measured area.}
 \label{fig:stubbed}
\end{figure*}

To see the relevance of the complex bands, first of all we focus our attention on the range of frequencies of the bandgap (yellow rectangular area of Figure \ref{fig:waveguide}a) of the perfect SC. Among the classical propagating and deaf bands,\cite{Vasseur05} we can observe evanescent connections (blue diamonds and green square lines). To analyze their properties, we analyze three different modes marked in the Figure with red squares. The mode 1 ($\Psi_1=0.484$, $K_1=1/2+0.049\imath$) belongs to a connecting evanescent band (green squared line) between two symmetric propagating bands. Notice that at the same frequency in the real part there is a deaf band (antisymmetric) that does not contribute to the propagation properties. The absolute value of the acoustic field obtained from the Fourier transform of the eigenvectors is in agreement with the predictions of the scattering problem shown in Figure \ref{fig:waveguide}b which are also experimentally measured using our experimental set up (Figure \ref{fig:waveguide}e). One can also observe the decay rate of the mode as it penetrates in the waveguide. This decay is related with the imaginary part of the Bloch vector predicted by EPWE, $Im(K_1)=0.049$. Fitting this decay to an exponential, $ae^{-bx}$,\cite{Romero10a, Romero10b}  we obtain $b=0.053\pm0.004$ with fairly good agreement with EPWE. On the other hand, the mode 2 ($\Psi_2=0.619$, $K_2=0.346$) belongs to a guiding (symmetric) band. Notice that for this frequency, the evanescent connection (green squared line) appears between two antisymmetric (deaf) bands therefore the evanescent mode cannot be excited. The MST predictions for this propagating mode is shown in Figure \ref{fig:waveguide}c and the experimental evidence is shown in Figure \ref{fig:waveguide}f.

We would like to pay special attention to mode 3 ($\Psi_3=0.731$, $K_3=0.3+0.21\imath$) in Figure \ref{fig:waveguide}a. This mode belongs to a complex band with real part inside the first Brillouin zone (blue diamond line). Notice that this evanescent symmetric band connects two symmetric (guided) bands. The maximum value of the lower band appears in the $\Gamma$ point (open black circle) whereas the minimum value of the upper band appears at the X point (open black circle), as a consequence the real part of the evanescent connection crosses the Brillouin zone.  This kind of bands is not observed by the classical $\omega(\vec{k})$ methods and it could considerably contribute to the propagation properties as we will see later. The acoustic field predicted by the Fourier transform of the eigenvectors again coincides with the predicted field using MST shown in Figure \ref{fig:waveguide}d. Notice the good agreement between the theoretical prediction and the experimental results in Figure \ref{fig:waveguide}g. On the other hand, for this case EPWE predicts a decay rate equal to $Im(K_3)=0.21$ which coincides with the fit from the MST prediction, $b=0.21\pm0.03$.

In order to observe the relevance of the evanescent modes and the accuracy of the EPWE predictions, we analyze the previous waveguide in which a scatterer is removed generating a stubbed SCW. Due to both the cavity and the local breaking of the symmetry produced by the stub, two effects can disturb the transmission properties of the system. On one hand if some resonance of the stub coincides with either a guiding mode or an evanescent wave, the resonance will be excited and the wave should be localized around the stub. On the other hand, antisymmetric modes can be excited due to the local breaking of the symmetry produced by the stub. The classical methods only predict the interaction of the guiding modes with this phenomena, but, as we have previously observed, evanescent waves can also appear in the system.

Left column of Figure \ref{fig:stubbed} shows the analysis for the frequencies of the resonances of the stub. The left and middle columns show the analysis for resonances at $\Psi=0.3812$ and $\Psi=0.605$ respectively. Each one of these frequencies coincide with a guiding band and as a consequence an hybridization of the resonance (flat band) and the guiding band appears (see left and middle panels in Figure \ref{fig:stubbed}a). The MST predictions in Figure \ref{fig:stubbed}b show the acoustic fields excited by an impinging plane wave. The acoustic field obtained using both MST and EPWE methods is fairly similar. The  Figure \ref{fig:stubbed}c shows the measured acoustic fields in the region marked with green squares in the Figure \ref{fig:stubbed}b. We can theoretically and experimentally observe how the guiding modes excite the resonances in the stub.

At this stage we would like to notice that the third resonance of the stub (right panels of the left column of Figure \ref{fig:stubbed}, $\Psi=0.72$ $K=0.29+0.20\imath$) coincides with an evanescent connection (blue diamond line in Figure \ref{fig:waveguide}a) predicted using EPWE. Thus, whereas using classical $\omega(\vec{k})$ methodologies no modes would be expected at this frequency, using the EPWE a coupling of the evanescent modes with the stub resonances appears. In the right panel of Figure \ref{fig:stubbed}a one can see that for this case no hybridization appears in the complex band structures. 
  As a consequence, one can theoretically (right panel of Figure \ref{fig:stubbed}b) and experimentally (right panel of Figure \ref{fig:stubbed}c) observe how the evanescent mode excites the resonance in the stub changing the transmission properties in good agreement with the EPWE predictions.

Recently the mode conversion in phononic crystals thin slabs have been observed.\cite{Chen08} Depending on the symmetry of the system with respect to the incident direction of the incident wave, shear-horizontal modes either couple or not with the Lamb wave modes. The coupling can be observed by a splitting in the lower-order symmetry band structure and by a transfer of the symmetry from one band to the other. Bavencoffe \cite{Bavencoffe09a} established a link between the attenuation of the ultrasonic wave observed in the case of a limited grating and the values of the imaginary part of the wave number in a stop band computed for an infinite grating. On the other hand Achaoui \textit{et al.} \cite{Achaoui10} have recently observed in phononic crystals with some freedom of anisotropy that when a band mostly polarized in-plane is close to a band mostly polarized out-of-plane, a phenomenon of repelling can occur and in some instances it introduces a local band gap. Moreover, the work shows that this interaction is accompanied by a transfer of the polarization state from one band to the other. These levels repulsion \cite{Dembowski01} which avoid crossing in the distribution of eigenvalues is well known within the physics community, however, there are only few works in the literature \cite{Wu04,Chen08,Achaoui10} analysing and discussing this phenomenon in two-dimensional phononic crystals. The stubbed SCW shown in this work seems to be a good example to observe the level repulsion. 

The complex band structure of the stubbed SCW at the frequencies of the levels repulsion are shown in Figure \ref{fig:stubbed}d. One can observe that between the two repelled guiding bands an evanescent connection appears. It is worth noting that the symmetry of lower (upper) guiding band changes from symmetric (antisymmetric) in the $\Gamma$ point to antisymmetric (symmetric) in the X point, appearing the evanescent connection between the maximum of the lower band and the minimum of the upper one. The evanescent connection presents a mixed symmetry showing the transfer of symmetry between the two repelled bands also observed in previous works \cite{Wu04, Chen08, Achaoui10}  but now explained as the presence of the evanescent connections. The acoustic fields excited by a plane wave at these frequencies are shown in Figure \ref{fig:stubbed}e. These fields are close to the predicted by the Fourier transform of the eigenvectors obtained using the EPWE for these frequencies and they are in good agreement with the experimental measurements corresponding to the area marked in green squares, shown in Figure \ref{fig:stubbed}f. The attenuation predicted between the two repelled bands is explained by the decay rate of the mode.

The complex band structures obtained using the EPWE ($k(\omega)$ method) show additional bands not revealed by the classical $\omega(\vec{k})$ methods being a fundamental tool to control the evanescent modes in periodic systems. In this work we have theoretically shown the interpretation of these evanescent additional bands with very good agreement with the experimental results. In the ranges of frequencies where a deaf band is traditionally predicted an evanescent mode with the excitable symmetry appears changing drastically the interpretation of the transmission properties of the system. On the other hand evanescent modes can excite cavity resonances in the stubbed SCW. We have interpreted, without loss of generality, the level repulsion between symmetric and antisymmetric bands in sonic crystals as the presence of an evanescent mode connecting both bands. These evanescent modes explain both the attenuation produced in this range of frequencies and the transfer of symmetry from one band to the other. Thus, the additional bands of EPWE should be considered to control the evanescent modes in the systems based on periodicity. The EPWE offers the possibility to known the spatial distribution of the field of the evanescent modes in periodic media, thus this can show the way to enhance the properties of the evanescent modes.

\begin{acknowledgments}
VRG and LMGR would like to thank the facilities provided by the IEMN UMR CNRS 8520. LMGR would like to thank the UPV for the grant PAID-00-11. VRG is grateful for the support contracts of the UPV CEI-01-11. This work was supported by MCI Secretar\'ia de Estado
de Investigaci\'on (Spanish government) and the FEDER funds, under grant MAT2009-09438.
\end{acknowledgments}

\end{document}